\documentclass[aps,pra,twocolumn,10pt,superscriptaddress,showpacs,amsmath,amssymb,nofootinbib]{revtex4-1}

\usepackage{graphicx}
\usepackage{dcolumn}
\usepackage{multirow}

\newcommand{\SNR}{\text{SNR}}
\newcommand{\AEE}{A_{\text{E1-E1}}}
\newcommand{\AEM}{A_{\text{E1-M1}}}
\newcommand{\AME}{A_{\text{M1-E1}}}
\newcommand{\AW}{A_{\text{W}}}
\newcommand{\AS}{A_{\text{S}}}
\newcommand{\HW}{H_{\text{W}}}
\newcommand{\HS}{H_{\text{S}}}
\newcommand{\chiS}{\chi_{\text{S}}}
\newcommand{\chiW}{\chi_{\text{W}}}
\newcommand{\bra}[1]{\langle #1|}
\newcommand{\ket}[1]{| #1\rangle}
\newcommand{\CG}[3]{\langle #1;#2|#3\rangle}
\newcommand{\ehat}{\mathbf{e}}
\newcommand{\xhat}{\ehat_x}
\newcommand{\yhat}{\ehat_y}

\newcommand{\evec}{\boldsymbol{\epsilon}}
\newcommand{\svec}{\boldsymbol{\sigma}}
\newcommand{\kvec}{\mathbf{k}}
\newcommand{\dvec}{\mathbf{d}}
\newcommand{\muvec}{\boldsymbol{\mu}}
\newcommand{\Efield}{\mathbf{E}}

\renewcommand{\Im}{\mathrm{Im}}

\newcommand{\kz}{k_{z}}

\begin{document}
\title{Atomic parity violation in $0\rightarrow0$ two-photon transitions}
\author{D.~R.~Dounas-Frazer}
\email{drdf@berkeley.edu}
\author{K.~Tsigutkin}
\author{D.~English}
\affiliation{Department of Physics, University of California at
Berkeley, Berkeley, California 94720-7300, USA}
\author{D.~Budker}
\affiliation{Department of Physics, University of California at
Berkeley, Berkeley, California 94720-7300, USA} \affiliation{Nuclear 
Science Division, Lawrence Berkeley National Laboratory, Berkeley, 
California 94720, USA}
\date{\today}


\begin{abstract}
We present a method for measuring atomic parity violation in the 
absence of static external electric and magnetic fields. Such 
measurements can be achieved by observing the interference of parity 
conserving and parity violating two-photon transition amplitudes 
between energy eigenstates of zero electronic angular momentum. 
General expressions for induced two-photon transition amplitudes are 
derived. The signal-to-noise ratio of a two-photon scheme using the 
$6s^2\;^1S_0\rightarrow 6s6p\;^3P_0$ transition in ytterbium is 
estimated. 
\end{abstract}

\maketitle

%
%
%
%
%
%
%
% =====================================================================
% Introduction
% =====================================================================
\section{Introduction}
High-precision measurements of parity violation in atoms, ions, and 
molecules constrain new physics beyond the Standard 
Model~\cite{ref:Ginges2004}. The processes that contribute to atomic 
parity violation (APV) are separated into two categories according 
to their dependence on nuclear spin~\cite{ref:Khriplovich1991}. The 
dominant contributions to APV usually come from nuclear 
spin-independent (NSI) processes, whereas nuclear spin-dependent 
(NSD) effects constitute small corrections. Measurements of NSI APV 
in cesium~\cite{ref:Wood1997} have led to precise evaluation of the 
weak charge of the nucleus~\cite{ref:Porsev2009}, and those of NSD 
APV to the first observation of the nuclear anapole 
moment~\cite{ref:Flambaum1997}. Future APV measurements may provide 
information about neutron distributions~\cite{ref:Horowitz2001}, and 
may reconcile the cesium anapole measurement with the limits placed 
on the anapole moment in thallium~\cite{ref:Vetter1995}.

APV experiments measure interference of a parity conserving 
transition amplitude with a parity violating one that is induced by 
the weak interaction~\cite{ref:Guena2005}. Such interference leads 
to circular birefringence in atomic vapors, which was employed in 
the earliest APV observations~\cite{ref:Barkov1978, 
ref:Bogdanov1980, ref:Macpherson1991, ref:Meekhof1995}. Other APV 
experiments, including the most accurate~\cite{ref:Wood1997} and the 
most recent~\cite{ref:Tsigutkin2009,*ref:Tsigutkin2010}, have 
employed the Stark-interference 
technique~\cite{ref:Bouchiat1974,*ref:Bouchiat1975}, a method that 
uses a static electric field to amplify an otherwise very small APV 
signal.

In addition to these well established Stark-interference techniques, 
there are various extensions: Light-shift measurements of amplitude 
interference have been proposed for various systems, including 
atoms~\cite{ref:Bouchiat2008}, single trapped 
ions~\cite{ref:Fortson1993, ref:Wansbeek2008}, two-ion entangled 
states~\cite{ref:Mandal2010}, and chiral 
molecules~\cite{ref:Harris1978, ref:Bargueno2009}. The potential 
advantages of using electromagnetically induced transparency to 
measure APV-induced circular dichroism in thallium have been 
investigated~\cite{ref:Cronin1998}. It has also been proposed to 
employ interference of a parity conserving two-photon transition 
with a parity violating single-photon transition in 
cesium~\cite{ref:GunawardenaPRL2007,*ref:GunawardenaPRA2007}.

All these methods rely on application of static external electric 
and magnetic fields to amplify and discriminate APV effects. 
Misalignments of applied fields introduce systematic uncertainties 
limiting the precision of APV measurements~\cite{ref:Tsigutkin2009}. 
In this work, we present a scheme for measuring NSI APV that 
replaces static electric and magnetic fields with optical fields 
that are easier to align. This scheme uses a two-photon transition 
between energy eigenstates with zero electronic angular momentum. 
Amplification of APV effects is achieved by interference of two 
transition amplitudes: a parity conserving amplitude describing 
electric-dipole-magnetic-dipole (E1-M1) transitions, and a parity 
violating E1-E1 amplitude. The APV signal can be discriminated from 
the large parity conserving background by manipulating properties of 
the light fields. A further advantage of this scheme is the ability 
to measure spurious electric and magnetic fields. This method, which 
we call the \emph{all-optical scheme (AOS)}, is applicable to a 
variety of atomic systems.

We consider an application of the AOS that takes advantage of the 
large NSI APV mixing of the $6s6p\;^1P_1$ and $5d6s\;^3D_1$ states 
observed in ytterbium~\cite{ref:Tsigutkin2009}. Precise measurements 
of this mixing in a chain of isotopes will provide important 
information about nuclear structure~\cite{ref:Brown2009}, and is a 
major goal of ongoing Stark-interference 
experiments~\cite{ref:Tsigutkin2009}. Systematic errors due to 
imperfections of applied fields pose a challenge for APV 
experiments, and cross-checks of present and future measurements are 
highly valuable. In the case of cesium, for instance, a cross check 
was provided by a stimulated-emission 
experiment~\cite{ref:Guena2003,*ref:Guena2005PRA}. To this end, we 
propose applying the AOS to the ytterbium two-photon 
($\lambda_1=399$~nm, $\lambda_2=1.28$~$\mu$m) 
$6s^2\;^1S_0\rightarrow 6s6p\;^3P_0$ transition to measure the 
parity-violating mixing of the intermediate $6s6p\;^1P_1$ state with 
the $5d6s\;^3D_1$ state.

From a formal point of view, the AOS is equivalent to measuring 
optical-rotation induced by APV on an M1 
transition~\cite{ref:BouchiatPC}. However, the AOS provides more 
field reversals compared to traditional optical-rotation 
experiments, thereby allowing for better discrimination of 
systematic effects from the APV signal. For the ytterbium system, a 
scheme for measuring APV-induced circular dichroism on the 
1.28~$\mu$m $6s6p\;^3P_0\rightarrow6s6p\;^1P_1$ has previously been 
proposed~\cite{ref:Kimball2001}. The AOS has two advantages over 
that proposal: the light fields are cw rather than pulsed, bypassing 
the challenges of achieving a high repetition rate; and, circular 
dichroism is measured by observing population of the metastable 
$6s6p\;^3P_0$ state, which allows for measurement in a region where 
detection conditions are easier to optimize.

%This paper is organized as follows. We describe the AOS in Sec. II, 
%and identify sources of systematic uncertainty in Sec. III. The 
%$6s^2\;^1S_0\rightarrow 6s6p\;^3P_0$ transition in ytterbium is 
%discussed in Sec. IV. A summary of the results is presented in Sec. 
%V. The two-photon transition amplitudes used in this work are 
%derived in the Appendices.

%
%
%
%
%
%
%
% =====================================================================
% AOS
% =====================================================================
\section{All-Optical Scheme}\label{sec:AOS}
We consider atoms illuminated by two light fields, with polarization 
vectors $\evec_j$, propagation vectors $\kvec_j$, and frequencies 
$\omega_j$, where $j=1,2$ is the light-field index. We denote the 
wavenumber $k_j\equiv|\kvec_j|=\omega_j/c$, the wavelength 
$\lambda_j=2\pi/k_j$, and the field intensity $\mathcal{I}_j$. The 
light fields drive two-photon transitions from initial state 
$\ket{i}$ to final state $\ket{f}$, separated in energy by 
$\omega_{fi}$. Throughout this work, we use atomic units: 
$\hbar=e=m_{\text{e}}=1/(4\pi\varepsilon_0)=1$. The transition rate 
on resonance ($\omega_1+\omega_2=\omega_{fi}$) 
is~\cite{ref:Faisal1987}:
\begin{equation}\label{eq:Rif}
R = (2\pi)^3 \alpha^2\mathcal{I}_1\mathcal{I}_2 |A|^2 
\frac{2}{\pi\,\Gamma},
\end{equation}
where $\alpha$ is the fine-structure constant, $A$ is the transition 
amplitude, and $\Gamma$ is the width of the transition. Energy 
eigenstates are represented as $\ket{i}=\ket{J_i M_i}$, and likewise 
for $\ket{f}$. Here $J_i$ and $M_i\in\{\pm J_i,\pm(J_i-1),\hdots\}$ 
are quantum numbers associated with the electronic angular momentum 
and its projection along the quantization axis, respectively. 

\begin{figure}[tb]
\includegraphics[width=\columnwidth]{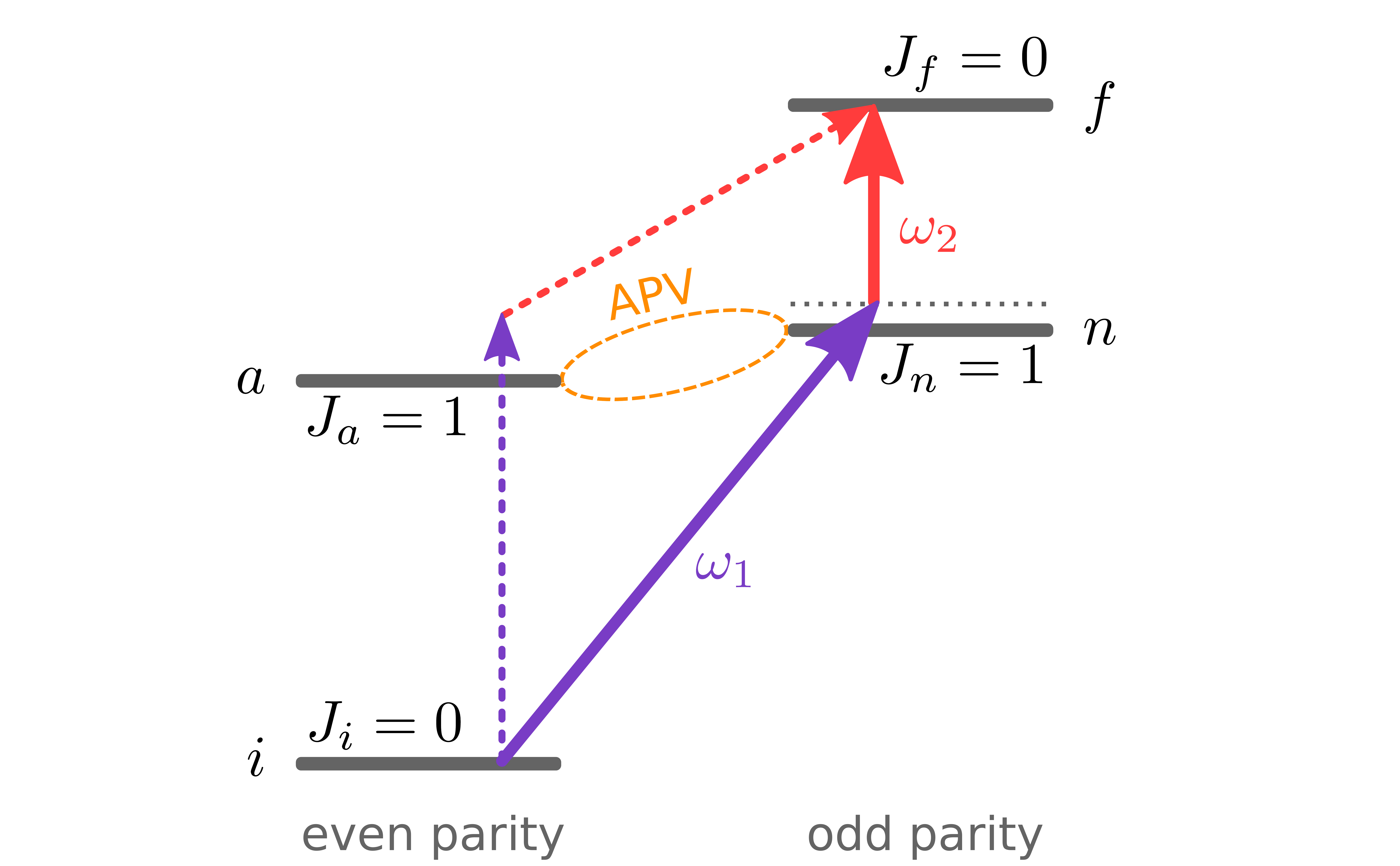}
\caption{\label{fig:EMI} (Color online) Energy levels and 
transitions relevant to the AOS. The dashed ellipse represents 
mixing of the states $\ket{n}$ and $\ket{a}$ due to the weak 
interaction. The dotted horizontal line represents the detuning of 
the light fields from the intermediate state. Thick, solid arrows 
and thin, dashed arrows illustrate dominant and suppressed 
excitation paths, respectively.}
\end{figure}

The AOS uses a two-photon transition from an initial state with 
$J_i=0$ to an opposite-parity final state with $J_f=0$. The 
transition is enhanced by the presence of an intermediate state 
$\ket{n}$ with $J_n=1$. The character of the two-photon transition 
depends on the magnitude of the detuning of the light fields from 
the one-photon resonances involving the intermediate 
state~\cite{ref:Ter-Mikaelyan1997}. When the detuning is small, the 
final state is populated by cascade excitation, that is, consecutive 
single-photon $i\rightarrow n$ and $n\rightarrow f$ transitions. We 
work in the opposite regime of large detuning [see 
condition~(\ref{eq:|delta|}) below]. In this regime, the excitation 
occurs via a pure two-photon transition and the population of the 
intermediate state is negligible.

The probability amplitude for the $i\rightarrow f$ transition has 
two contributions: one from a parity conserving E1-M1 transition, 
and another from a parity violating E1-E1 transition. The E1-E1 
transition is induced by mixing of the intermediate state\footnote{
Alternatively, the E1-E1 transition could be induced by mixing of 
the final state $\ket{f}$ with a nearby opposite-parity $J=0$ state. 
Such a scheme is equivalent to the AOS, and shares the key features 
of the AOS presented in this work.
} $\ket{n}$ with opposite-parity $J=1$ states via the weak 
interaction. We assume that this mixing is dominated by a single 
nearby state $\ket{a}$ with $J_a=1$ (Fig.~\ref{fig:EMI}). The 
proximity of $\ket{a}$ to $\ket{n}$ leads to an M1-E1 excitation 
path for the $i\rightarrow f$ transition that uses the intermediate 
state $\ket{a}$ rather than $\ket{n}$. We incorporate the amplitude 
of this path into the expression for the E1-M1 amplitude below.

In the absence of stray fields, the amplitude for a two-photon 
$J_i=0\rightarrow J_f=0$ transition is~(Appendix~A):
\begin{equation}\label{eq:A00}
A = \AEM + \AW,
\end{equation}
where
\begin{equation}
\AEM =[\mathcal{M}(\omega_1)\hat{\kvec}_2 - 
\mathcal{M}(\omega_2)\hat{\kvec}_1] 
\cdot(\evec_1\times\evec_2)\label{eq:AEM}
\end{equation}
and
\begin{equation}
\AW = i[\zeta(\omega_1)+\zeta(\omega_2)] 
(\evec_1\cdot\evec_2)\label{eq:AW}
\end{equation}
are the amplitudes corresponding to the E1-M1 and weak interaction 
induced E1-E1 transitions\footnote{
The E1-M1 amplitude is a product of the matrix elements of the 
operators describing E1 and M1 single-photon transitions. Because 
the E1 (M1) operator is odd (even) under spatial inversion, the 
E1-M1 amplitude is a pseudoscalar. A similar argument shows that the 
E1-E1 amplitude is a normal scalar.
}. The quantities $\mathcal{M}(\omega_j)$ and $\zeta(\omega_j)$ are
\begin{equation}
\mathcal{M}(\omega_j) = \frac{1}{3}\left( 
\frac{\mu_{fn}\,d_{ni}}{\omega_{ni}-\omega_j} 
+\frac{d_{fa}\,\mu_{ai}}{\omega_{ai}-\omega_j} \right) \label{eq:M}
\end{equation}
and
\begin{equation}
\zeta(\omega_j)= 
\frac{1}{3}\frac{d_{fa}\,\Omega_{an}\,d_{ni}}{\omega_{na}}\left( 
\frac{1}{\omega_{ni}-\omega_j}- 
\frac{1}{\omega_{ai}-\omega_j}\right) ,\label{eq:zeta} 
\end{equation}
where $\mu_{fn}$ and $d_{ni}$ are the reduced matrix elements of the 
magnetic- and electric-dipole moments, respectively. Here 
$\omega_{na}$ is the energy splitting of states $\ket{a}$ and 
$\ket{n}$, and $\Omega_{an}$ is the magnitude of the reduced matrix 
element of the NSI APV Hamiltonian $\HW$ [see Eq.~(\ref{eq:chiweak}) 
in the Appendix].

The two terms in Eqs.~(\ref{eq:M}, \ref{eq:zeta}) correspond to the 
two different excitation paths for the transition. When states 
$\ket{n}$ and $\ket{a}$ are perfectly degenerate ($\omega_{na}=0$), 
the induced E1-E1 paths interfere destructively and the parity 
violating amplitude vanishes. We limit our discussion to atomic 
systems for which $\omega_{na}$ is sufficiently large that one path 
is dominant. We assume that $\Delta=\omega_1-\omega_{ni}$ is much 
smaller than all other detunings from the intermediate states 
$\ket{n}$ and $\ket{a}$. In this case, only
\begin{equation}
\mathcal{M}(\omega_1)\approx\mathcal{M}\equiv \frac{1}{3} 
\frac{\mu_{fn}\,d_{ni}}{\Delta}
\end{equation}
and
\begin{equation}
\zeta(\omega_1)\approx\zeta\equiv \frac{1}{3}
\frac{d_{fa}\,\Omega_{an}\,d_{ni}}{\omega_{na}\Delta}
\end{equation}
contribute significantly to Eqs.~(\ref{eq:AEM}, \ref{eq:AW}). 
Because $\mathcal{M}$ and $\zeta$ have the same complex 
phase~\cite{ref:Khriplovich1991}, the relative phase between $\AEM$ 
and $\AW$ is determined by the field geometry. Hereafter, we assume 
that $\mathcal{M}$ and $\zeta$ are real parameters since their 
common phase is arbitrary.
%In practice, the sums in Eqs.~(\ref{eq:M}, \ref{eq:zeta}) 
%are dominated by terms from a single $\ket{n}$ and $\ket{a}$.

The goal of the AOS is to observe interference of amplitudes $\AW$ 
and $\AEM$ in the rate $R$. As Eqs.~(\ref{eq:Rif}-\ref{eq:AW}) show, 
$R$ consists of a large parity conserving term proportional to 
$\mathcal{M}^2$, a small parity violating term (the interference 
term) proportional to $\mathcal{M}\zeta$, and a negligibly small 
term on the order of $\zeta^2$. The interference term is 
proportional to a pseudoscalar quantity that depends only on the 
field geometry, the \emph{rotational invariant}:
\begin{equation}\label{eq:rotinv00}
\Im\{(\evec_1\cdot\evec_2)^{\ast}[(\evec_1\times\evec_2)\cdot\kvec_2]\},
\end{equation}
which is odd under parity reversal (P-odd) and even under time 
reversal (T-even)\footnote{ 
The rotational invariant presented in Eq.~(\ref{eq:rotinv00}) is not 
symmetric under photon exchange because we have neglected the terms 
proportional to $\mathcal{M}(\omega_2)$ and $\zeta(\omega_2)$ in 
Eqs.~(\ref{eq:AEM},~\ref{eq:AW}). When these terms are included, $R$ 
has two interference terms whose sum is exchange symmetric. In the 
case of degenerate photons ($\omega_1=\omega_2$), the sum of the 
interference terms is proportional to the exchange symmetric 
rotational invariant 
$\Im\{(\evec_1\cdot\evec_2)^{\ast}[(\evec_1\times\evec_2) 
\cdot(\hat{\kvec}_1-\hat{\kvec}_2)]\}$.
}. %Here $\Im\{\circ\}$ is the imaginary part of the term in the 
%curly braces, and $(\evec_1\cdot\evec_2)^{\ast}$ is the complex 
%conjugate of $(\evec_1\cdot\evec_2)$.
The time reversal invariance of expression (\ref{eq:rotinv00}) can 
be understood in the following way. Time reversal requires that 
initial and final states be interchanged, which corresponds to 
complex conjugation. The imaginary part of a complex number is 
therefore T-odd. Since the photon momentum $\kvec_2$ is also T-odd, 
the rotational invariant (\ref{eq:rotinv00}) is T-even.

Restrictions on the geometry of the light fields are inferred from 
the form of the rotational invariant. The rotational invariant--and 
hence the interference term--vanishes for plane polarized light 
beams. One way to achieve a nonzero rotational invariant is to 
choose circular polarization for the second beam: 
$\evec_2=\svec_{\pm}$, where we have assumed $\kvec_2$ lies along 
the $z$ axis. For arbitrary polarization $\evec_1=a_+\svec_+ + 
a_-\svec_-$ of the first beam, conservation of angular momentum 
requires that only the polarization component $a_{\mp}$ contributes 
to the excitation process. Thus the rotational 
invariant~(\ref{eq:rotinv00}) reduces to $\pm|a_{\mp}|^2\kz$, where 
$\kz=+1$ ($\kz=-1$) when $\kvec_2$ is aligned (anti-aligned) with 
the $z$-axis.

For two collinear circularly polarized beams of light oriented along 
the $z$-axis, the transition rate is
\begin{equation}\label{eq:R00}
R\propto |A|^2 = \mathcal{M}^2\pm 2\kz\mathcal{M}\zeta ,
\end{equation}
where the positive (negative) sign in Eq.~(\ref{eq:R00}) is taken 
when beam 2 has $\svec_+$ ($\svec_-$) polarization. The interference 
term is discriminated from the total transition rate either by 
reversing the direction of the propagation vector $\kvec_2$, 
%rotating the polarization angle 
%$\theta$ by $\pi/2$, switching $\evec_2$ from linear ($\varphi=0$) to 
%circular ($\varphi=\pi/4$) polarization,
or by reversing the sense of rotation of the circularly polarized 
light fields. The asymmetry 
$2\zeta/\mathcal{M}=2(d_{fa}/\mu_{fn})(\Omega_{an}/\omega_{na})$ is 
obtained by dividing the difference of rates upon a reversal by 
their sum.

We propose to measure the transition rate by probing the population 
of the final state $\ket{f}$, and assume that the following 
conditions are met: (i) detuning of the light fields from the 
intermediate state is large enough to realize a pure two-photon 
transition, and (ii) light intensities are low enough that the 
transition is not saturated. A pure two-photon transition is 
achieved when~\cite{ref:Faisal1987,ref:Ter-Mikaelyan1997}
\begin{equation}\label{eq:|delta|}
|\Delta|\gg\Omega_0,
\end{equation}
where 
\begin{equation}\label{eq:Omega0}
\Omega_0 = \sqrt{\frac{8\pi\alpha}{3} \Big(d_{ni}^2\,\mathcal{I}_1+ 
\mu_{fn}^2\,\mathcal{I}_2\Big)}
\end{equation}
is the interaction energy. When condition (\ref{eq:|delta|}) is 
satisfied, the system reduces to a two-level system consisting of 
the initial and final states coupled by an effective optical field. 
Saturation effects can be ignored when the pumping rate $R$ is much 
smaller than the relaxation rate $\Gamma'$ of the final state, that 
is, when
\begin{equation}\label{eq:I1I2}
\mathcal{I}_1\mathcal{I}_2\ll\left[ \frac{3
}{\pi\alpha}\frac{\Delta}{d_{ni}\,\mu_{fn}}\right]^2\Gamma\Gamma'.
\end{equation}
In this regime, the population of the final state is proportional to 
the rate given by Eq.~(\ref{eq:Rif}).

The shot-noise limited sensitivity of the AOS is estimated as 
follows. The probe signal is proportional to the number of excited 
atoms. The number of excited atoms associated with the parity 
violating part of the transition rate is
\begin{equation}
N = [(4\pi\alpha)^2 \mathcal{I}_1 \mathcal{I}_2 (2\mathcal{M} 
\zeta)/ \Gamma] N_i t,
\end{equation}
where $t$ is the effective integration time of the measurement and 
$N_i$ is the total number of atoms initially in state $\ket{i}$. The 
measurement noise is given by $\delta N= \sqrt{N'}$, where 
\begin{equation}
N'=[(4\pi\alpha)^2 \mathcal{I}_1 \mathcal{I}_2 \mathcal{M}^2/ 
\Gamma]N_it
\end{equation}
is the number of excited atoms associated with the parity conserving 
part of the transition rate. The signal-to-noise ratio~(SNR) of the 
probe signal is $N/\delta N$, or
\begin{equation}\label{eq:SNR00}
\SNR=\frac{8\,\pi\alpha}{3}\,
\frac{d_{fa}\,\Omega_{an}\,d_{ni}}{\omega_{na}|\Delta|} \sqrt{ 
\frac{\mathcal{I}_1\,\mathcal{I}_2\,N_i\,t}{\Gamma} }.
\end{equation}
As expected, the SNR increases for large $\Omega_{an}$ and high 
light intensities. Although purely statistical shot-noise dominated 
SNR does not contain $\mu_{fn}$, this amplitude is still an 
important parameter in practice due to conditions~(\ref{eq:|delta|}, 
\ref{eq:I1I2}). Large $\mu_{fn}$ leads to small APV asymmetry which 
requires better control over experimental parameters (see 
Sec.~\ref{sec:systematics}). In the opposite case of small 
$\mu_{fn}$, an observable signal requires high light intensities 
which may pose a technical challenge.

Although we have focused on a ladder-type three-level atomic system 
(Fig.~\ref{fig:EMI}), the discussion presented here holds for 
lamda-type systems (Fig.~\ref{fig:ybAOS}) as well. However, the 
following modifications must be made: In a lambda-type system, the 
polarization of the absorbed photon $\evec_2$ is replaced by the 
polarization of the photon emitted by stimulated emission. 
Consequently, conservation of energy requires that the two-photon 
resonance condition above Eq.~(\ref{eq:Rif}) becomes 
$\omega_1-\omega_2=\omega_{fi}$, and conservation of angular 
momentum requires that both circularly polarized beams have the same 
sense of rotation. Then the positive (negative) sign is taken in 
Eq.~(\ref{eq:R00}) when $\evec_1=\evec_2=\sigma_{-(+)}$.

%
%
%
%
%
%
%
% =====================================================================
% Systematics
% =====================================================================
\section{Parasitic sources of asymmetry}\label{sec:systematics}
Systematic effects may also contribute to the asymmetry and mask the 
APV signal. In this section, we discuss three potential sources of 
such parasitic asymmetry: imperfections of applied optical fields; 
Stark interference due to stray electric fields, and; shifts of the 
intermediate state energy induced by external fields.

\subsection{Optical field imperfections}
To understand the effects of imperfections in the optical fields, we 
relax the assumptions of collinear light beams and perfectly 
circular polarization. Beam misalignment is characterized by the 
angle $\theta\ll 1$ between $\kvec_2$ and $\kvec_1$. We choose the 
$z$-axis to lie along $\kvec_2$ so that $\theta$ is the polar angle 
of $\kvec_1$. Deviations from circular polarization are 
characterized by the parameters $\eta_j\ll1$. Here $|\varphi_j| = 
\pi/4-\eta_j$ is the magnitude of the ellipticity\footnote{
The polarization $\evec_j$ of arbitrarily polarized light is 
parameterized in terms of the polarization angle $\vartheta_j$ and 
the ellipticity $\varphi_j$ in the following way: $\evec_j = 
(\cos\vartheta_j\cos\varphi_j- i 
\sin\vartheta_j\sin\varphi_j)\xhat+(\sin\vartheta_j\cos\varphi_j+ 
i\cos\vartheta_j\sin\varphi_j)\yhat$. Linearly, circularly, and 
elliptically polarized light are described by $\varphi_j=0$, 
$|\varphi_j|=\pi/4$ and $0<|\varphi_j|<\pi/4$, respectively. The 
sense of rotation is determined by the relative sign of the real and 
imaginary parts of $\evec_j$: $\evec_j=\svec_\pm$ when 
$\varphi_j=\pm\pi/4$.
} of the $j$th beam.

Field imperfections lead to additional parity conserving 
terms\footnote{
Beam misalignment is further characterized by the azimuthal angle 
$\phi$ of $\kvec_1$. Taking this into account, the transition rate 
becomes $R\rightarrow 
R-\mathcal{M}^2[\theta^2/2+\eta_1^2+\eta_2^2+2\eta_1\eta_2\cos(2\vartheta-2\phi)]$, 
where $\vartheta=\vartheta_2-\vartheta_1$ is the relative 
polarization angle of the light fields. In the text we treat the 
case of maximal correction, that is, $ \vartheta=\phi$.
} in $R$ on the order of 
$\mathcal{M}^2[\theta^2/2+(\eta_1+\eta_2)^2]$. Although these 
corrections to $R$ do not mimic APV, they nevertheless contribute to 
the asymmetry if they change upon reversal. To simplify our 
analysis, we assume that changes in$\theta$ and $\eta_j$ between 
reversals are on the order of $\theta$ and $\eta_j$ for any 
particular reversal. We further assume that $\eta_1$ and $\eta_2$ 
are the same order of magnitude: $\eta_1\simeq\eta_2\equiv\eta$. 
Then spurious ellipticity and beam misalignment give rise to a 
parasitic asymmetry on the order of $\theta^2 + 2\eta^2$, and may 
mask the APV signal if they are large. Asymmetry due to field 
imperfections is negligible compared to APV asymmetry when
\begin{equation}
\theta^2 \ll 2\zeta/\mathcal{M} \quad\text{and}\quad \eta^2 \ll 
\zeta/\mathcal{M}.
\end{equation}

%
%
%
%
%
%
%
% =====================================================================
% Systematics
% =====================================================================
\subsection{Stark interference}
The derivation of Eq.~(\ref{eq:R00}) assumes the absence of external 
fields. Here we relax this assumption and discuss the uncertainty in 
the AOS that arises due to spurious electric fields. In the presence 
of a static electric field $\Efield$, Stark mixing of $\ket{n}$ and 
$\ket{a}$ induces an E1-E1 transition between the initial and final 
states. The Stark-induced transition amplitude is (Appendix~A):
\begin{equation}
\AS=i\xi [\Efield\cdot(\evec_1\times\evec_2)]\label{eq:AS0},
\end{equation}
where
\begin{equation}\label{eq:beta}
\xi=\frac{1}{3\sqrt{6}}\, 
\frac{d_{fa}\,d_{an}\,d_{ni}}{\omega_{na}\Delta}.
\end{equation}
After including the effects of stray fields and misalignments, the 
transition rate~(\ref{eq:R00}) becomes
\begin{align}
R &\propto \mathcal{M}^2+\xi^2E_zE_1 \pm \kz\big(\mathcal{M}\xi 
E_\perp + 2\mathcal{M}\zeta \big), \label{eq:R'}
\end{align}
where $E_z$ is the $z$ component of the electric field, 
$E_1=\Efield\cdot\hat{\kvec}_1\approx E_z$, 
$E_{\perp}=|\mathbf{E}\cdot(\hat{\kvec}_1\times\hat{\kvec}_2)|\lesssim\theta 
|\Efield|$, and only terms linear in $\theta$ and $\eta_j$ are 
presented. The terms in the parentheses represent the combined 
effects of APV and Stark interference. These interference terms 
exhibit the same behavior under reversals of the light fields. This 
means that there is a contribution to the asymmetry on the order of 
$\theta \xi E/\mathcal{M}$, where $E$ is the magnitude of 
$|\Efield|$. APV asymmetry dominates over asymmetry due to Stark 
interference when
\begin{equation}\label{eq:condition}
E \ll 2\zeta/(\theta \xi).
\end{equation}

\subsection{Energy shifts of the intermediate state}
The transition rate must be further modified to account for light 
shifts and effects of stray magnetic fields. We consider energy 
shifts of the form $M_n^2\delta_2+M_n\delta_1$, where $M_n$ is the 
magnetic quantum number of $\ket{n}$. The parameter $\delta_2$ is 
due to tensor shifts caused by the light fields or dc electric 
fields. We neglect quadratic Zeeman shifts that arise in the 
presence of transverse magnetic fields. In general, many levels may 
contribute to light shifts of the intermediate state. We approximate 
light shifts by their contributions from the initial and final 
states. Then the light shifts are approximately equal to 
$\Omega_0^2/(4\Delta)$, where $\Omega_0$ is given by 
Eq.~(\ref{eq:Omega0}). When the dc polarizability of the 
intermediate state is dominated by Stark mixing of $\ket{n}$ and 
$\ket{a}$, dc Stark shifts of that state are on the order of 
$d_{an}^2E^2/(2\omega_{na})$. Then
\begin{equation}
\delta_2\approx \Omega_0^2/(4\Delta)+ d_{an}^2E^2/(2\omega_{na}).
\end{equation}
On the other hand, the parameter $\delta_1$ is due to vector light 
shifts and the dc Zeeman effect. Vector light shifts change sign 
upon reversal of circular polarization 
($\svec_{\pm}\rightarrow\svec_{\mp}$). In this case,
\begin{equation}
\delta_1 \approx \pm\Omega_0^2/(4\Delta) + g\mu_0B,
\end{equation}
where $g$ is the Land\'e factor of the intermediate state, $B$ is 
the magnitude of the stray magnetic field, and $g\mu_0B$ is the 
order of magnitude of the Zeeman shift.

The corrected E1-M1 and Stark-induced E1-E1 transition amplitudes 
are obtained by expanding the energy denominators in 
Eqs.~(\ref{eq:M}, \ref{eq:beta}) to first order in $\delta_{1,2}$. 
Corrections to the weak interaction induced 
amplitude~(\ref{eq:zeta}) are neglected here. The transition rate is
\begin{align}
R&\propto\mathcal{M}^2\big[1+2(\delta_2\pm\delta_1)/\Delta\big] + 
\xi^2E_1E_z -\nonumber\\ &\quad 
-2\xi^2E_z^2(\delta_2\pm\delta_1)/\Delta \pm 
 \kz\big(\mathcal{M}\xi E_\perp + 
2\mathcal{M}\zeta \big),\label{eq:R''}
\end{align}
where summation over the magnetic sublevels of $\ket{n}$ has been 
taken into account. Only the Stark interference term $\mathcal{M}\xi 
E_\perp$ has the same signature as APV. The Stark and Zeeman shift 
corrections can be discriminated from the other terms in 
Eq.~(\ref{eq:R''}) by changing the sign of $\Delta$, and can be 
discriminated from each other by reversing the sense of rotation of 
the circularly polarized light fields. Thus stray electric and 
magnetic fields can be measured by alternating the sign of the 
detuning of the light fields from the intermediate state.

%
%
%
%
%
%
%
% =====================================================================
% Application of AOS
% =====================================================================
\section{Applications to ytterbium}~\label{sec:Yb}

\begin{figure}[tb]
\includegraphics[width=\columnwidth]{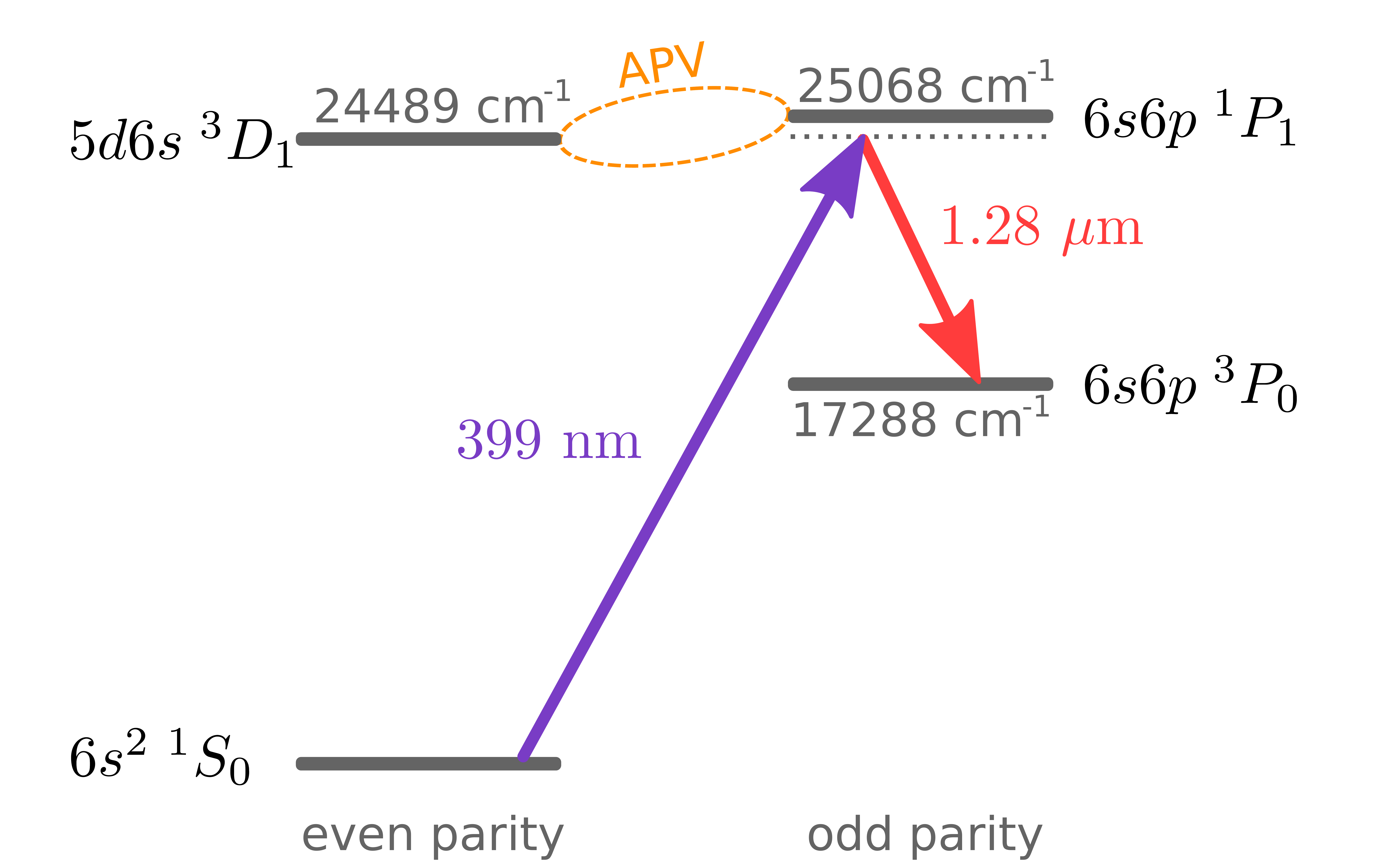}
\caption{\label{fig:ybAOS} (Color online) Application of AOS to 
ytterbium.}
\end{figure}

\begin{table}
\caption{\label{tab:AOS} Atomic data for application of AOS to 
ytterbium. Here $a_0$ and $\mu_0$ are the Bohr radius and magneton.}
\begin{ruledtabular}
\renewcommand{\arraystretch}{1.2}
\begin{tabular}{l*{4}{>{$}l<{$}}}
&\multicolumn{1}{c}{Transition ($1\rightarrow 2$)} & 
\multicolumn{1}{c}{$d_{21}/(ea_0)$} & 
\multicolumn{1}{c}{$\mu_{21}/\mu_0$}
\\\hline
\multirow{3}{*}{E1}
 & 6s^2~~~\,^1S_0 \rightarrow6s6p\;^1P_1  & 4.1\footnotemark[1] \\
 & 6s6p~\,^3P_0 \rightarrow5d6s\;^3D_1  & 2.6\footnotemark[2] \\
 & 5d6s~^3D_1 \rightarrow6s6p\;^1P_1  & 0.27\footnotemark[2]\\ \hline
\multirow{2}{*}{M1}
 & 6s^2~~~\,^1S_0 \rightarrow5d6s\;^3D_1 & & 1.33\footnotemark[3]\times10^{-4}
\\
 & 6s6p~\,^3P_0 \rightarrow6s6p\;^1P_1 & & 0.13\footnotemark[4]\\
\end{tabular}
\footnotetext[1]{Ref.~\cite{ref:Blagoev1994}}
\footnotetext[2]{Ref.~\cite{ref:Porsev1999}}
\footnotetext[3]{Ref.~\cite{ref:Stalnaker2002}} 
\footnotetext[4]{Ref.~\cite{ref:Kimball2001}}
\end{ruledtabular}
\end{table}

We now turn our attention to the two-photon $6s^2\;^1S_0\rightarrow$ 
$6s6p\;^3P_0$ transition in ytterbium. This transition can be driven 
by two light fields of wavelengths $\lambda_1=399$~nm and 
$\lambda_2=1.28$~$\mu$m, which are nearly resonant with transitions 
involving the intermediate $6s6p\;^1P_1$ state 
(Fig.~\ref{fig:ybAOS}). The parity violating E1-E1 transition is 
induced by the mixing of the $6s6p\;^1P_1$ and $5d6s\;^3D_1$ states 
due to the weak interaction. This mixing arises because 
$6s6p\;^1P_1$ has a large admixture of the configuration 
$5d6p$~\cite{ref:DeMille1995, ref:Dzuba2011}. The parameter 
describing mixing of $6s6p\;^1P_1$ and $5d6s\;^3D_1$ was measured to 
be $\Omega_{an}/\omega_{na} = 
6\times10^{-10}$~\cite{ref:Tsigutkin2009}. Other essential atomic 
parameters are given in Table~\ref{tab:AOS}. The ytterbium system is 
characterized by the asymmetry 
$2\zeta/\mathcal{M}\approx6\times10^{-6}$, which is more than an 
order of magnitude larger than asymmetries measured in 
optical-rotation experiments in bismuth, lead, and 
thallium~\cite{ref:Ginges2004}.

For concreteness, we consider a hypothetical experiment using an 
atomic beam similar to that of Ref.~\cite{ref:Tsigutkin2009}: 
characteristic thermal speed $3\times10^4$~cm/s, density 
$2\times10^9$~cm$^{-3}$, radius 1~cm. The atomic beam intersects two 
overlapping, collinear laser beams where atoms interact with the 399 
nm and 1.28 $\mu$m light and undergo the $6s^2\;^1S_0\rightarrow 
6s6p\;^3P_0$ transition. High light powers--which are necessary to 
achieve a large SNR--can be realized using a unidirectional ring 
cavity. The transition rate can be measured by probing the 
population of the metastable $6s6p\;^3P_0$ state using the detection 
method described in Ref.~\cite{ref:Tsigutkin2009}.

To estimate the SNR, we choose light parameters that satisfy 
conditions (\ref{eq:|delta|}, \ref{eq:I1I2}). The laser beams have a 
Guassian profile with a characteristic radius of 2~mm, and the 
frequencies are detuned from the intermediate state by $\Delta= 
2\pi\times800$~MHz, about 30 times larger than the width of the 
intermediate state. In the interaction region, the metastable state 
acquires a radiative decay rate on the order of 
$\Gamma'=[\Omega_0^2/(4\Delta^2)]\tau^{-1}=2\pi\times30$~kHz, where 
$\tau=5.68$~ns is the lifetime of 
$6s6p\;^1P_1$~\cite{ref:Blagoev1994}. The width of the 
$6s^2\;^1S_0\rightarrow 6s6p\;^3P_0$ transition is dominated by the 
transit-broadened linewidth\footnote{
The transit-broadened width of a one-photon transition is
$\Gamma_0\approx2.4v/r$~\cite{ref:Demtroder2003}, where $v$ is the 
atomic velocity and $r$ is the $1/e$ radius of the light electric 
field profile. The matrix element for a two-photon transition is 
proportional to the product of the two light electric fields, and 
hence to the product of their Gaussian 
profiles~\cite{ref:Faisal1987}. Therefore, the transit-broadened 
width is $\Gamma=\sqrt{2}\Gamma_0$ when the beam profiles are 
identical.
} $\Gamma=2\pi\times90$~kHz. For light powers of 10~W at 1.28-$\mu$m 
and 10~mW at 399-nm, Eq.~(\ref{eq:SNR00}) gives 
$\SNR\approx2\sqrt{t(\text{s})}$. Based on these estimates, a 
one-hour measurement may achieve better than 1\% statistical 
uncertainty in determination of parity violation. %Because we have 
%not taken into account the time required to perform field reversals 
%when computing the SNR, our result is an over-estimate. The SNR may 
%be further reduced by inefficiencies of the detection apparatus and 
%Doppler broadening of the spectral line.

Parasitic asymmetry due to field imperfections and stray electric 
fields can be controlled by aligning the laser beams over a large 
distance. In order to limit asymmetry due to beam misalignment to 
less than 1\% of the APV asymmetry, the angle between the nominally 
collinear laser beams must be controlled to better than one-tenth of 
a beam radius of transverse displacement over a distance of one 
meter, which corresponds to a beam misalingment of 0.01$^\circ$. 
Similarly, the deviation from circular polarization must also be 
smaller than about 0.01$^\circ$. In this case, the systematic 
uncertainty due to Stark interference effects is below 1\% for 
electric fields smaller than 8 V/cm. In 
Ref.~\cite{ref:Tsigutkin2009}, stray electric fields were measured 
to be on the order of 1~V/cm. In that experiment, stray electric 
fields are partially attributed to charge buildup on surfaces of 
electrodes and coils that are used to generate external static 
electric and magnetic fields. For the AOS, the absence of such 
surfaces will likely result in even smaller stray fields.

%In principle, the magnitude of stray electric fields can be obtained 
%by measuring the static Stark shift using reversals of $\Delta$ and 
%$\sigma$ (Sec.~\ref{sec:systematics}). However, the shifts induced 
%by 1-V/cm fields lead to corrections to the transition rate on the 
%order of $d_{an}^2E^2/(4\omega_{an}\Delta)\approx10^{-12}$ which are 
%likely too small to be measured.

It is important to consider other mechanisms for population of the 
metastable $6s6p\;^3P_0$ state, causing background and noise. The 
metastable state may be populated by multiphoton processes involving 
highly excited states, or by molecular processes in the presence of 
dimers or other molecular impurities in the atomic beam. These 
detrimental effects will contribute to a background that dependends 
on $\Delta$, compromising the search for stray 
fields~\cite{ref:BouchiatPC}. We note that no evidence of molecular 
impurities has been seen in the Yb APV epxeriments up to 
date~\cite{ref:Tsigutkin2009}.

To better understand the feasibility of the proposed experiment, we 
compare the predicted SNR of the two-photon AOS to the observed SNR 
of the one-photon Stark-interference 
experiment~\cite{ref:Tsigutkin2009}. This comparison is especially 
relevant since both techniques employ the same method for probing 
the population of the metastable $6s6p\;^3P_0$ state in ytterbium. 
The shot-noise limited SNR in Ref.~\cite{ref:Tsigutkin2009} was 
demonstrated to be $2\sqrt{t(\text{s})}$. However, the 
Stark-interference experiment is not currently shot-noise limited; 
the measurement uncertainty is determined by systematic effects due 
to imperfections of applied fields. Because the AOS has similar 
projected statistical sensitivity and possibly better control of 
systematics compared to the Stark-interference experiment, the AOS 
is an attractive candidate for future APV measurements in ytterbium.

%
%
%
%
%
%
%
% =====================================================================
% Summary and discussion
% =====================================================================
\section{Summary and discussion}
A scheme for measuring APV using interference of parity conserving 
E1-M1 and parity violating E1-E1 two-photon transition amplitudes 
was presented. The AOS allows for observations of NSI APV in the 
absence of external static electric and magnetic fields. This method 
measures the rate of a transition between two energy eigenstates 
with zero total electronic angular momentum. General expressions for 
the two-photon transition rate and SNR were derived. Because the AOS 
uses optical fields rather than static electric and magnetic fields, 
systematic effects due to field misalignments are easier to minimize 
in the AOS than in ongoing APV 
measurements~\cite{ref:Tsigutkin2009}.

To demonstrate the feasibility of the AOS, we estimated the SNR of 
the $6s^2\;^1S_0\rightarrow6s6p\;^3P_0$ transition in ytterbium 
($\lambda_1=399$~nm, $\lambda_2=1.28$~$\mu$m). Our estimate of the 
SNR suggests that this system is a promising candidate for a 
cross-check of recent APV measurements in 
ytterbium~\cite{ref:Tsigutkin2009}, and future measurements of APV 
in a chain of isotopes. While we considered atoms with zero nuclear 
spin, the AOS could also be applied to isotopes with nonzero nuclear 
spin provided that the detuning $\Delta$ is much larger than the 
hyperfine splitting of the intermediate state.

Another candidate for the AOS is the ladder-type 
$4f^66s^2\;^7F_0\rightarrow 4f^66s6p\;^7F_0$ transition 
($\lambda_1=639$~nm, $\lambda_2=1.56$~$\mu$m) in samarium. The E1-E1 
transition is induced by mixing of the opposite-parity states 
$4f^65d6s\;^7G_1$ and $4f^66s6p\;^7G_1$ due to the weak interaction. 
The APV effect in samarium is expected to be of the same order of 
magnitude as that observed in ytterbium~\cite{ref:Rochester1999}. A 
version of the AOS that uses photons of the same frequency has been 
previously suggested in a proposal for a search for APV using the 1 
$\mu$m $1s2p\;^3P_0\rightarrow1s2s\;^1S_0$ transition in uranium 
ions~\cite{ref:Schafer1989}. However, the uranium-ion experiment is 
not currently feasible because the required laser intensity is on 
the order of $10^{21}$~W/cm$^2$.

%
%
%
%
%
%
%
% =====================================================================
% Acknowledgements
% =====================================================================
\section{Acknowledgements}
The authors acknowledge helpful discussions with M.~A.~Bouchiat, 
A.~Cing\"oz, M.~Kozlov, N.~A.~Leefer, and J.~Stalnaker. This work 
has been supported by NSF.

%
%
%
%
%
%
%
% =====================================================================
% APPENDIX
% =====================================================================
\appendix

\section{Two-photon transition amplitudes}\label{appendix}
In this section, we derive amplitudes for E1-M1 and induced E1-E1 
$J_i=0\rightarrow J_f=0$ transitions. We 
use the following convention for the Wigner-Eckart theorem (WET). 
Let $T_k$ be an irreducible tensor of rank $k$ with spherical 
components $T_{kq}$ for $q\in\{0,\pm1,\hdots,\pm k\}$. Then the WET 
is~\cite{ref:Sobelman1992}
\begin{equation}\label{eq:WET}
\bra{J_2M_2}T_{kq}\ket{J_1M_1} = 
(J_2||T_k||J_1)\frac{\CG{J_1M_1}{kq}{J_2M_2}}{\sqrt{2J_2+1}},
\end{equation}
where $(J_2||T_k||J_1)$ is the reduced matrix element of $T_k$ and 
$\CG{J_1M_1}{kq}{J_2M_2}$ is a Clebsch-Gordan coefficient.

The amplitude for the E1-M1 transition is~\cite{ref:Faisal1987}
\begin{equation}\label{eq:AEM_start}
\begin{split}
\AEM = {} & \AEM(1,2) + \AEM(2,1)\\
& +\AME(1,2) + \AME(2,1),
\end{split}
\end{equation}
where
\begin{equation}\label{eq:AAEM}
\AEM(j,j') = 
\bra{f}(\hat{\kvec}_{j'}\times\evec_{j'})\cdot\muvec 
\,\frac{\ket{n}\bra{n}}{\omega_{ni}-\omega_j}\,
\evec_j\cdot\dvec\ket{i},
\end{equation}
and
\begin{equation}\label{eq:AAME}
\AME(j,j')=\bra{f}\evec_{j'}\cdot\dvec
\,\frac{\ket{a}\bra{a}}{\omega_{ai}-\omega_j}\,
(\hat{\kvec}_j\times\evec_j)\cdot\muvec\ket{i},
\end{equation}
for $j,j'=1,2$. Here $\muvec$ and $\dvec$ are the magnetic- and electric-dipole 
moments of the atom, and summation over the magnetic sublevels of 
the intermediate states $\ket{n}$ and $\ket{a}$ is implied.

E1-E1 transitions are induced by mixing of the states $\ket{n}$ and 
$\ket{a}$ due to the weak interaction and, in the presence of a 
static electric field, the Stark effect. The perturbed states 
$\ket{n}+\chi\ket{a}$ and $\ket{a}-\chi^{\ast}\ket{n}$ act as 
intermediate states for the two paths that contribute to the induced 
E1-E1 amplitude. Here $\chi$ is a small dimensionless parameter that 
depends on the details of the perturbing Hamiltonian. The amplitude 
for the induced E1-E1 transition is~\cite{ref:Faisal1987}
\begin{equation}\label{eq:AEE_start}
\AEE = \AEE(1,2)+\AEE(2,1),
\end{equation}
where
\begin{equation}
\begin{split}
\AEE(j,j') = \bra{f}\evec_{j'}\cdot\dvec\left[
\frac{\chi\ket{a}\bra{n}}{\omega_{ni}-\omega_j}-
\frac{\ket{a}\bra{n}\chi}{\omega_{ai}-\omega_j}\right]\evec_j\cdot\dvec\ket{i}.
\end{split}
\end{equation}
Like for Eqs.~(\ref{eq:AAEM}, \ref{eq:AAME}), summation over the 
magnetic sublevels of states $\ket{n}$ and $\ket{a}$ is implied.

When the mixing of $\ket{n}$ and $\ket{a}$ is due to the weak 
interaction alone, the perturbation parameter is given by 
$\chi=\chiW$ where
\begin{equation}\label{eq:chiweak}
\chiW = \frac{\bra{a}\HW\ket{n}}{\omega_{na}} = 
\frac{i}{\sqrt{3}}\frac{\Omega_{an}}{\omega_{na}},
\end{equation}
for $J_a=J_n$ and $M_a=M_n$. Here $\Omega_{an}$ is a real parameter 
related to the reduced matrix element of $\HW$ by 
$(J_a||\HW||J_n)=i\Omega_{an}$. The factor of $i$ preserves time 
reversal invariance~\cite{ref:Khriplovich1991}. 

In the presence of a static electric field $\Efield$, the 
perturbation parameter becomes $\chi = \chiW + \chiS$, where 
$\chiW$ is given by Eq.~(\ref{eq:chiweak}) and
\begin{equation}
\chiS = \frac{\bra{a}\HS\ket{n}}{\omega_{na}} = 
-\frac{d_{an}E_{q}^{\ast}}{\omega_{na}} 
\frac{\CG{J_nM_n}{1q}{J_aM_a}}{\sqrt{2J_a+1}},
\end{equation}
where $d_{an}$ is the reduced matrix element of the electric-dipole 
operator. Here $\HS=-\dvec\cdot\Efield$ is the Stark Hamiltonian, 
$E_q$ is the $q$th spherical component of $\Efield$, 
$E^{\ast}_q=(-1)^qE_{-q}$, and $q=M_a-M_n$. In this case, 
$\AEE=\AW+\AS$, where $\AW\propto\chiW$ and $\AS\propto\chiS$ are 
the amplitudes of the transitions induced by the weak interaction 
and Stark effect, respectively.

For a general $J_i\rightarrow J_f$ transition, the Stark-induced 
E1-E1 amplitude may have contributions from each of the irreducible 
tensors that can be formed by the three vectors $\evec_1$, 
$\evec_2$, and $\Efield$. There are seven such tensors: one of rank 
0, three of rank 1, two of rank 2, and one of rank 3. However, for a 
$J_i=0\rightarrow J_f=0$ transition, only the rank-0 tensor 
contributes. This tensor is~\cite{ref:Varshalovich1988}
\begin{align}
T_{00}&=\sum_{\lambda,q}
\CG{1\lambda}{1q}{00}\{\evec_1\otimes\evec_2\}_{1\lambda}E_q\nonumber\\
&=-\frac{i}{\sqrt{6}}\Efield\cdot(\evec_1\times\evec_2),
\end{align}
where
\begin{align}
\{\evec_1\otimes\evec_2\}_{1\lambda} &= 
\sum_{\mu,\nu}\CG{1\mu}{1\nu}{1\lambda} 
\epsilon_{1\mu}\epsilon_{2\nu}\nonumber\\
&=\frac{i}{\sqrt{2}}(\evec_1\times\evec_2)_\lambda
\end{align}
is the irreducible tensor of rank 1 formed by $\evec_1$ and 
$\evec_2$. Here $\lambda,\;q,\;\mu,\;\nu=0,\;\pm1$ are the spherical 
components of $\{\evec_1\otimes\evec_2\}_{1}$, $\Efield$, $\evec_1$, 
and $\evec_2$, respectively. The Stark-induced transition amplitude 
is
\begin{equation}
\AS=i[\xi(\omega_1)-\xi(\omega_2)] 
[\Efield\cdot(\evec_1\times\evec_2)].\label{eq:ASapp}
\end{equation}
The coefficient $\xi(\omega_j)$ can be expressed in terms of reduced 
electric-dipole matrix elements by applying the WET to 
Eq.~(\ref{eq:AEE_start}) with $\chi=\chiS$ and comparing the result 
to Eq.~(\ref{eq:ASapp}). This procedure yields
\begin{equation}\label{eq:betaapp}
\xi(\omega_j)=\frac{1}{3\sqrt{6}}\,
\frac{d_{fa}\,d_{an}\,d_{ni}}{\omega_{na}}
\left(\frac{1}{\omega_{ni}-\omega_j}-
\frac{1}{\omega_{ai}-\omega_j}\right).
\end{equation}
The Stark effect may also cause the final state $\ket{f}$ to mix 
with nearby opposite-parity $J=1$ states. In this case, 
Eq.~(\ref{eq:ASapp}) is still valid, but Eq.~(\ref{eq:betaapp}) must 
be modified to account for additional admixtures of states.

Expressions~(\ref{eq:AEM}, \ref{eq:AW}) for the amplitudes of the 
E1-M1 and weak interaction-induced E1-E1 transitions are derived by 
direct application of the WET to Eqs.~(\ref{eq:AEM_start}, 
\ref{eq:AEE_start}). The Stark-induced E1-E1 amplitude in 
Eq.~(\ref{eq:ASapp}) reduces to expression~(\ref{eq:AS0}) when 
$|\Delta|\ll|\omega_{na}|$ and
 $|\Delta|\ll|\omega_2-\omega_{ni}|$.

\end{document}